\documentclass[amsmath,amssymb]{revtex4}

\usepackage{graphicx}
 \usepackage{xcolor}
    
\begin{document}
\title{Heat exchange for oscillator strongly coupled to thermal bath}
\author{Alex V. Plyukhin}
\email{aplyukhin@anselm.edu}
 \affiliation{ Department of Mathematics,
Saint Anselm College, Manchester, New Hampshire 03102, USA 
}

\date{\today}

\begin{abstract}

The heat exchange fluctuation theorem (XFT) by  
Jarzynski and W\'ojcik [Phys. Rev. Lett. 92, 230602 (2004)] addresses  the setting 
where two systems with different temperatures are brought in thermal contact 
at time $t=0$ and then disconnected at later time $\tau$.
The theorem asserts that  the probability of  an anomalous heat flux (from cold to hot), while nonzero, 
is exponentially smaller than the probability of the corresponding normal flux (from hot to cold). 
As a result, the average heat flux is always normal. In that way, the theorem demonstrates how 
irreversible heat transfer, observed on the macroscopic scale, emerges from the underlying reversible dynamics. 
The XFT was proved under the assumption that the coupling work required to connect and 
then disconnect the systems is small compared to the change of the internal energies of the systems.
That condition is often valid for macroscopic systems, 
but may be violated for microscopic ones.
We examine the validity of the XFT's assumption for a specific model of the Caldeira-Leggett type,  where one system 
is a single classical harmonic oscillator and the other is a thermal  bath comprised of a large number of oscillators. 
The coupling between the system and the bath, which is bilinear, is instantaneously turned on at $t=0$ 
and off at $t=\tau$. For that model, we found that the assumption of the XFT 
can be satisfied only for a rather restricted range of parameters. In general, the work involved in the process
is not negligible and the energy exchange may be anomalous 
in the  sense that the internal energy of the system, which is initially hotter than the bath,  may further increase. 

\end{abstract}

\maketitle
\section{Introduction}

Consider two systems $A$ and $B$,
initially disconnected and prepared in thermal equilibrium at temperatures $T_A$ and $T_B$, respectively.  
At time $t=0$ the systems are placed in thermal contact, exchange heat during the time interval $(0,\tau)$, and then disconnected at time $t=\tau$.
For this settings, Jarzynski and W\'ojcik  derived a fluctuation theorem for heat exchange, christening it as XFT~\cite{JW}. The theorem  states  that the probability distribution $f(q)$ of heat    
$q$ transferred from $A$ to $B$ satisfies the relation 
\begin{eqnarray}
    \frac{f(+q)}{f(-q)}=\exp\left(
    q\,\Delta \beta\right),
    \label{xft}
\end{eqnarray}
where $\Delta \beta=T_B^{-1}-T_A^{-1}$. This relation 
quantifies the probabilities of the normal (from hot to cold) and the anomalous (from cold to hot) heat transfer. It shows that while anomalously directed transfer  is possible as a fluctuation, the average transfer is always normally directed. 
Suppose $T_A>T_B$, $\Delta\beta>0$, and the heat $q$ is defined to be positive when it transferred from $A$ to $B$.  Then Eq. (\ref{xft}) predicts that the normal transfer of heat  $q>0$ from hot $A$ to cold $B$ is exponentially more probable than the anomalous transfer of heat $-q<0$, i.e. of heat $q>0$  from cold $B$ to hot $A$. As a result, the average heat $Q=\langle q\rangle $ transferred from hot $A$ to cold $B$ is positive,
\begin{eqnarray}
    Q=\int_{-\infty}^\infty f(q)\,q\, dq=
    \int_0^\infty f(q)\,q\, \left(1-e^{-q \Delta \beta}\right)\, dq>0.
    \label{heat_transfer}
\end{eqnarray}
For the isothermal process, $T_A=T_B$, $\Delta \beta=0$, the average transferred heat is zero, $Q=0$. These results hold for any values of the process duration $\tau$ and account for the second law of thermodynamics in the form that the average heat on any time scale flows from hot to cold. 

The XFT~(\ref{xft}) was derived in the weak coupling approximation, i.e. under the assumption that  the (coupling) work $W$, required to connect and later disconnect $A$ and $B$, is negligible. In that case, the (microscopic) heat $q$ transferred from $A$ to $B$ is naturally defined as the change of the energy $u_B(t)$ of system $B$,
\begin{eqnarray}
    q =\Delta u_B=u_B(\tau)-u_B(0).
    \label{Q_def0}
\end{eqnarray}
It also equals the negative change of the energy $u_A(t)$ of system $A$, 
\begin{eqnarray}
    q=-\Delta u_A=-u_A(0)+u_A(\tau).
    \label{Q_aux1}
\end{eqnarray}
The microscopic energy changes of the two systems may have arbitrary but opposite signs (or both be zero),  
\begin{eqnarray}
    \Delta u_B=-\Delta u_A=q.
    \label{Q_aux2}
\end{eqnarray}
Taking the average one gets
\begin{eqnarray}
    \Delta U_B=-\Delta U_A=Q,
    \label{Q_aux3}
\end{eqnarray}
where  
$U_A=\langle u_A\rangle$ and $U_B=\langle u_B\rangle$ are the internal energies of the systems, and the average heat $Q$ transferred from $A$ to $B$ equals to the change 
of the internal energy of system  $B$,
\begin{eqnarray}
    Q=\Delta U_B.
    \label{Q_def}
\end{eqnarray}
As noted above, the fluctuating energy changes $\Delta u_A$ and $\Delta u_B$ may be of arbitrary (but opposite) sign. In contrast, 
the signs of average energy changes $\Delta U_A$, $\Delta U_B$ are not arbitrary, but determined by the sign of the temperature difference. For $T_A>T_B$ Eq. (\ref{heat_transfer}) gives
$Q=\Delta U_B>0$ and  Eq. (\ref{Q_aux3}) prescribes 
$\Delta U_A<0$. 

Statistical properties of the energy fluxes, using setups similar to that of the XFT,  have been studied experimentally, numerically, and theoretically in many works, see  ~\cite{Sancho,Ciliberto,Ciliberto2,Berut,Naert,Landi,Pal} and references therein. Although the findings 
are commonly consistent with the XFT, they  are often limited to the asymptotic case 
$\tau\to\infty$. Another often adopted limit is that the two involved systems are both macroscopic and can be treated as thermal baths with infinite heat capacities, whose temperatures 
are fixed and do not change during the process. For such case,
Gomez-Marin and Sancho~\cite{Sancho} addressed the question of how to extend the XFT
beyond the weak coupling limit. For 
a specific model (where two baths are connected by a Brownian transducer), they found 
a heat fluctuation theorem in the form similar to Eq. (\ref{xft}), but with $\Delta\beta$ replaced by  a more complicated function of the two temperatures.  The sign of that function, however,
coincides with that of $\Delta\beta$. Then, the evaluation similar to Eq. (\ref{heat_transfer}) shows again that the average heat flows from hot system $A$ to cold system $B$. 
It was found in~\cite{Sancho} that  the heat flux can be reversed  (to flow from cold to hot) but only when, in addition to the coupling work,  an external work (not related  to connection and disconnection of the systems) is applied. 
Quantum thermodynamic models with variable system-bath couplings were recently studies in~\cite{Wiedmann,Carrega,An}.

One difficulty of extending the XFT beyond the weak coupling limit is that the definition of heat may be  ambiguous~\cite{JW,Sancho,Landi,Pal,An}.
Assuming that  the combined system $AB$ is thermally isolated, the coupling work equals the change of the internal energy of the combined system $AB$, $W=\Delta U_{AB}=U_{AB}(\tau)-U_{AB}(0)$. For the XFT setting,  at $t=0$ and $t=\tau$ the two systems are disconnected, therefore the energy of the combined system equals merely the sum of those for systems $A$ and $B$,
$\Delta U_{AB}=\Delta U_A+\Delta U_B$ and does not include the interaction energy. Therefore we can write  
\begin{eqnarray}
    W=\Delta U_A+\Delta U_B.
    \label{aux0}
\end{eqnarray}
Let us emphasize that for the XFT setup this relation is exact and does not assume that the interaction between $A$ and $B$ is weak. 
Instead of Eq. (\ref{Q_aux3}), $\Delta U_B=-\Delta U_A$, which holds only for $W=0$, we  have in general 
\begin{eqnarray}
    \Delta U_B=-\Delta U_A+W.
\label{aux2}
\end{eqnarray}
If we still adopt expression (\ref{Q_def}) as the definition of the average heat $Q$ involved in the process, $Q=\Delta U_B$, i.e. as the energy received by system $B$, 
then  Eq. (\ref{aux2}) can be re-written as
\begin{eqnarray}
    Q=-\Delta U_A+W. 
    \label{law1}
\end{eqnarray}
This relation has the inviting form of the first law 
for system $A$, which may be useful in some situations~\cite{Aurell}. However, in general the interpretation of Eq. (\ref{law1}) as the first law for $A$
ought to be taken with reservations. 
First, 
one may argue that definition~(\ref{Q_def}), $Q=\Delta U_B$, is inconsistent with the notion of heat as the form of energy which is distinct from work. For the given process the work  does contributes to the energy change of system $B$, and therefore to $Q$. 
Second, the coupling work $W$ in Eq. (\ref{law1}) is not the work on $A$ alone, but on the combined system $AB$.

In general, the extension of thermodynamic concepts to  
 microscopic scale is a subtle problem which was addressed in recent years in many works~\cite{Gelin,Seifert,TH,Jar2,Esposito,PP}. In the XFT setting, the problem is simpler than  in general because Eq. (\ref{aux0}) does not involves the interaction energy but  
only bare energy changes of individual systems, $\Delta U_A$ and $\Delta U_B$.  Together with the coupling work $W$, those quantities are sufficient to characterize  XFT-like processes, and the concept of heat is actually unnecessary. Thus, instead of the standard set of variables
\begin{eqnarray}
    S_1=(W,\,Q,\,\Delta U),
    \label{set1}
\end{eqnarray}
for a system of interest only, $A$ or $B$, 
for the XFT protocol it is more natural to use the set 
\begin{eqnarray}
    S_2=(W,\,\Delta U_A,\,\Delta U_B),
    \label{set2}
\end{eqnarray}
which characterizes both systems involved in the process. 
While the variables of set $S_1$ are related by the first law $Q=W-\Delta U$, the variables of set $S_2$ are connected by Eq. (\ref{aux0}).
The latter can be referred to and interpreted as the first law for the XFT protocol. 
Using set $S_2$, one can describe  a broader range of processes than with set $S_1$.
One can consider $S_1$ as a subset of $S_2$ for which the additional constrain $\Delta U_A=-\Delta U_B=-Q$ holds.

One expects that the standard description of heat exchange in terms of set $S_1$ is still relevant and the XFT provides a good approximation if the coupling work is much smaller than the internal energy changes
\begin{eqnarray}
    |W|\ll |\Delta U_A|, \qquad |W|\ll |\Delta U_B|.
    \label{cond}
\end{eqnarray} 
While these conditions appear reasonable and intuitive for macroscopic systems, for small systems they are 
difficult to justify in general because 
the process depends on many parameters. 
The goal of this paper is to consider the validity of conditions (\ref{cond}) for a specific model which  allows an explicit evaluation of $W$, $\Delta U_A$, $\Delta U_B$.

To this end, 
we exploit the standard  Caldeira-Leggett model~\cite{CL,Weiss} modified for the XFT setup.
Although the XFT can be formulated in both classical and quantum cases, we restrict the discussion to the former.   
System $A$, referred to as ``the system", is represented by a classical oscillator with initial temperature $T_0$, and system $B$ is a thermal bath at temperature $T$ represented by a large set of oscillators. 
We denote the internal energy of the system and the  bath as $U$ and $U_b$, respectively.
We do not attempt to define heat involved in the XFT process. Instead, we 
describe the process in terms of the variables 
\begin{eqnarray}
    (W,\,\Delta U,\, \Delta U_b)
\end{eqnarray}
connected by the first law (\ref{aux0})
\begin{eqnarray}
    W=\Delta U+\Delta U_b.
    \label{law11}
\end{eqnarray}
For the presented model one can evaluate $W$ and $\Delta U$ independently,  
then $\Delta U_b$ is determined by Eq. (\ref{law11}). 
We found that conditions (\ref{cond}) may be satisfied only for a limited range of parameters. In general,  the values of $W$, $\Delta U$, $\Delta U_b$ are comparable, so the assumption of the XFT is not satisfied.
As a result, the average energy transfer 
is not determined by the temperatures only, as 
both thermodynamics and the XFT imply. 
In particular, for finite process duration $\tau$ the internal energy of a hotter system may increase. Such behavior does not contradict the second law because the process is not spontaneous but involves non-zero  coupling work $W$. We found that the latter may be  positive or negative, depending on parameters of the model. 

\section{Model}
We adopt the Caldeira-Leggett model~\cite{CL,Weiss} modified by allowing the system-bath coupling to be a function of time. Our system is  
a classical harmonic oscillator and the bath is represented by $N\gg 1$ independent oscillators.
The bare Hamiltonians are  
\begin{eqnarray}
H_s=\frac{p^2}{2m}+\frac{1}{2}\,m\,\omega^2 q^2
\end{eqnarray}
for the system, and 
\begin{eqnarray}
H_b=\sum_{i=1}^N\left(
\frac{p_i^2}{2m_i}+\frac{1}{2}\,m_i\,\omega_i^2 q_i^2
\right),
\label{H_b}
\end{eqnarray}
for the bath. 
The  system-bath interaction, when it is turned on, has the form
\begin{eqnarray}
H_{int}=-q\sum_{i=1}^N c_i q_i +
\frac{q^2}{2}\sum_{i=1}^N \frac{c_i^2}{m_i\omega_i^2}. 
\label{H_c}
\end{eqnarray}
Here $c_i$ are coupling coefficients, and 
the first term represents a bilinear coupling  of coordinates of the system  ($q$) and the bath ($q_i$). The second term, which we shall also write as
\begin{eqnarray}
V_c=\frac{1}{2}\,m\,\omega_c^2q^2, \quad 
\omega_c^2=
\frac{1}{m}\sum_{i=1}^N \frac{c_i^2}{m_i\omega_i^2},
\label{Omega_c}
\end{eqnarray}
is quadratic in the system's coordinate and 
does not depend on bath variables. It 
is required to compensate the driving effect of the bilinear coupling~\cite{CL,Weiss}.
The characteristic  frequency $\omega_c$, defined by Eq. (\ref{Omega_c}), is related  
to parameters of the Langevin dynamics  of the system, as will be discussed below.

The total Hamiltonian of the combined system  is 
\begin{eqnarray}
    H=H_s+H_b+\gamma(t)\,H_{int},
    \label{H_total}
\end{eqnarray}
where the coupling parameter $\gamma(t)$  evolves according to the protocol 
\begin{eqnarray}
    \gamma(t)=\begin{cases}
     0, & t<0 \\
     1, & 0\le t< \tau\\
     0, & t\ge\tau.
   \end{cases}
   \label{protocol}
\end{eqnarray}
The protocol  describes a process when the system and the bath, disconnected for $t<0$, are instantaneously coupled at $t=0$ and then instantaneously disconnected  at $t=\tau$.
The Hamiltonians of the uncoupled ($\gamma=0$) and coupled ($\gamma=1$) configurations are 
\begin{eqnarray}
H_0=H_s+H_b, \qquad H_1=H_s+H_b+H_{int}.\label{H01}
\end{eqnarray}
We assume that for $t<0$, when the system and the bath are uncoupled, they
are prepared in thermal equilibrium with temperatures 
$T_0$ and $T$, respectively. At $t=0^-$, i. e. just before the connection, 
the distributions of the system and of the bath are 
\begin{eqnarray}
\rho_s=Z_s^{-1}\,e^{-H_s/T_0}, \quad \rho_b=Z_b^{-1}\, e^{-H_b/T},
\end{eqnarray}
respectively,
and the combined system is described by the distribution  
\begin{eqnarray}
\rho_0=\rho_s\,\rho_b.
\label{rho0}
\end{eqnarray}
We assume that the preparation of the combined system in distribution $\rho_0$ is arranged at negligible energy cost.
The sudden coupling at $t=0$ brings the system and the bath temporally out of equilibrium.
For $0<t<\tau$ the combined system evolves according to the Hamiltonian dynamics with Hamiltonian $H_1$.

For the XFT protocol,
the average work $W$ involved
in the process is the sum of the 
the average work $W_0$ to connect the system and bath at $t=0$ and the average work $W_1(\tau)$ to disconnect them at $t=\tau$,
\begin{eqnarray}
    W(\tau)=W_0+W_1(\tau).
\end{eqnarray}
At $t=0$,
the Hamiltonian of the combined system is instantaneously changed from $H_0$ to $H_1$. 
Therefore, the average work to connect the system and the bath  is 
\begin{eqnarray}
  W_0=\big\langle H_1-H_0\big\rangle=\big\langle H_{int}\big\rangle.
\label{Wc0}
\end{eqnarray}
Here and below  the brackets $\langle \cdots\rangle$ 
denote averaging over the system and bath variables  with
distribution $\rho_0$, given by Eq. (\ref{rho0}), which is the distribution for the combined system at $t=0^-$.
Taking into account that  
$\langle q\,q_i\rangle=0$
and $\langle q^2\rangle=T_0/(m\, \omega^2)$, we  get
\begin{eqnarray}
W_0=\langle V_c\rangle=
\frac{1}{2}\, m\,\omega_c^2\langle q^2\rangle=\frac{T_0}{2}\,\left(\frac{\omega_c}{\omega}\right)^2,
\label{Wc}
\end{eqnarray}
where the compensation potential $V_c$ and  frequency $\omega_c$ are defined by Eq. (\ref{Omega_c}). 
The work $W_0$ is positive and represents the energy cost to increase the oscillator frequency from $\omega$ to $\omega+\omega_c$ instantaneously.
Note that if such an increase is arranged not instantaneously but quasistatically, the corresponding work $W_0$ can be shown to be zero.

The disconnection work $W_1$ corresponds to the instantaneous switch at $t=\tau$ of the Hamiltonian of the combined system from $H_1$ to $H_0$ and equals
\begin{eqnarray}
W_1(\tau)=\big\langle H_0(\tau)-H_1(\tau)\big\rangle=
  -\big\langle H_{int}(\tau)\big\rangle,
\end{eqnarray}
or 
\begin{eqnarray}
W_1(\tau)=\sum_{i=1}^N c_i\, \langle q(\tau)\, q_i(\tau)\rangle -
\frac{1}{2}\, m\,\omega_c^2\, \langle q^2(\tau)\rangle.
\label{W1}
\end{eqnarray}
Although this expression involves coordinates of both the system and the bath, it can be expressed in  terms  of coordinate and velocity of the system only, see Eq. (\ref{aux234}) below. Therefore, in order to evaluate $W_1$, and the total work $W=W_0+W_1$, is is sufficient to solve the equation of motion of the system only, eliminating bath degrees of freedom with the standard procedure outlined in the next section.

\section{General results}

 For $0<t<\tau$ 
the dynamics of the combined system
is described by Hamiltonian $H_1$ which corresponds
to the standard Caldeira-Leggett model.
Integrating the equations of motion of the bath 
\begin{eqnarray}
m_i\,\ddot q_i(t)=-m_i\,\omega_i^2\,q_i(t)+c_i\, q(t) 
\label{eq_bath}
\end{eqnarray}
and substituting the result into the equation of motion for the system 
\begin{eqnarray}
    m\,\ddot q(t)=-m\,\omega^2 q(t)+\sum_{i} c_i\, q_i(t) -m\,\omega_c^2\,q(t)
    \label{eq_system}
\end{eqnarray}
one can present  the latter   
in the form 
of the generalized Langevin equation~\cite{Weiss,Hanggi} 
\begin{eqnarray}
    \ddot q(t)=-\omega^2q(t)-\int_0^t\! K(t-\tau)\, \dot q(\tau)\,d\tau+\frac{1}{m}\,\xi(t)
    -\,K(t)\,q(0),
\label{gle}
\end{eqnarray}
where the term $\xi(t)$ can be interpreted as stationary and zero-centered noise connected to 
the dissipation kernel $K(t)$ by the fluctuation-dissipation relation,
\begin{eqnarray}
    \langle \xi(t)\rangle=0, \quad \langle \xi(t)\,\xi(t')\rangle=m\,T\, K(t-t'). 
    \label{fdr}
\end{eqnarray}
Strictly speaking,  in these expressions the averages are taken over initial variables of the bath only with 
distribution $\rho_b=Z_b^{-1} e^{-H_b/T}$. However, since $\xi(t)$ depends only on the initial values of the bath's variables~\cite{Weiss,Hanggi}, 
the averages can be replaced by those over the total distribution $\rho_0=\rho_s\,\rho_b$.

The explicit expression for the kernel $K(t)$ in Eq. (\ref{gle}) is~\cite{Weiss,Hanggi}
\begin{eqnarray}
    K(t)=\frac{1}{m}\sum_{i=1}^N \frac{c_i^2}{m_i\,\omega_i^2}\,\cos(\omega_i t).
    \label{K}
\end{eqnarray}
Comparing with  Eq. (\ref{Omega_c}), 
one observes  a relation between the  initial value of the dissipation 
kernel and the characteristic frequency $\omega_c$:
\begin{eqnarray}
    K(0)=\frac{1}{m}\sum_{i=1}^N \frac{c_i^2}{m_i\,\omega_i^2}=
    \omega_c^2.
    \label{K0}
\end{eqnarray}

The last term in the Langevin equation (\ref{gle}) is often referred to as the initial slip. Its presence is 
due to the specific initial conditions adopted in our model when the system and the bath  
at $t=0$ are not in equilibrium with each other~\cite{Weiss,Hanggi}.

From Eq. (\ref{eq_system}) one observes
\begin{eqnarray}
    \sum_{i=1}^Nc_i\,q_i(t)=m\,[\ddot q(t)+(\omega^2+\omega_c^2)\,q(t)].
    \label{aux23}
\end{eqnarray}
Substituting this to Eq. (\ref{W1}) allows one to eliminate the bath's variables and to express the disconnection work $W_1$ in terms of the system's variables only,
\begin{eqnarray}
 W_1(\tau)=m\,\langle \ddot q(\tau)\, q(\tau)\rangle
+m\,\omega^2\left[
1+\left(\frac{\omega_*}{\omega}\right)^2
\right]
\langle q^2(\tau)\rangle.
\label{aux233}
\end{eqnarray}
Here and below, to get rid of the coefficient $1/2$, we use instead of $\omega_c$ the characteristic frequency
\begin{eqnarray}
    \omega_*=\omega_c/\sqrt{2}.
\end{eqnarray}

We prefer to  present the above result for $W_1$
in terms of the second moments of the system's coordinate and velocity. 
Since $\ddot q q=\frac{d}{dt} (\dot q q) - \dot q^2=\frac{1}{2}\,\frac{d^2}{dt^2} q^2-\dot q^2$, 
one finds
\begin{eqnarray}
 W_1(\tau)=\frac{m}{2}\,\frac{d^2}{dt^2} \langle q^2(\tau)\rangle
    -m\,\langle \dot q^2(\tau)\rangle
+m\,\omega^2\left[
1+\left(\frac{\omega_*}{\omega}\right)^2
\right]
\langle q^2(\tau)\rangle.
\label{aux234}
\end{eqnarray}

Recall that the connection work $W_0$ is given by Eq. (\ref{Wc}), so 
for the total average work  $W=W_0+W_1$ we obtain
\begin{eqnarray}
W(\tau)=
T_0\,\left(\frac{\omega_*}{\omega}\right)^2+
    \frac{m}{2}\,\frac{d^2}{dt^2} \langle q^2(\tau)\rangle
    -m\,\langle \dot q^2(\tau)\rangle
+m\,\omega^2\left[
1+\left(\frac{\omega_*}{\omega}\right)^2
\right]
\langle q^2(\tau)\rangle. 
\label{W_total}
\end{eqnarray}

To proceed, one needs to solve the Langevin equation (\ref{gle}) and 
evaluate the moments $\langle q^2(\tau)\rangle$ and  $\langle \dot q^2(\tau)\rangle$.
The standard steps (see the Appendix) lead to the following results
\begin{eqnarray}
    \langle q^2(t)\rangle&=&\frac{T_0}{m\,\omega^2}\,R^2(t)+
    \frac{1}{m}\left(T_0-T\right)
    \,G^2(t)+
    \frac{T}{m\,\omega^2}\, [1-S^2(t)],\nonumber\\
    \langle \dot q^2(t)\rangle&=&\frac{T_0}{m\,\omega^2}\,\dot R^2(t)+
\frac{T_0}{m}\,R^2(t)+
    \frac{T}{m}\, [1-R^2(t)-\omega^2 G^2(t)].
    \label{moments}
\end{eqnarray}
Here the Green's function $G(t)$ is determined in the Laplace domain as
\begin{eqnarray}
    \tilde G(s)=\int_0^\infty\!\! e^{-st}G(t)\, dt=\frac{1}{s^2+s\tilde K(s) +\omega^2},
    \label{Green_def}
\end{eqnarray}
and the other two Green's functions $R(t)$ and $S(t)$ are related to $G(t)$ by equations
\begin{eqnarray}
    \dot G(t)=R(t), \quad \dot S(t)=-\omega^2 G(t).
    \label{green_derivatives}
\end{eqnarray}
The initial values of the Green's functions are  
\begin{eqnarray}
    G(0)=0, \quad R(0)=S(0)=1,
    \label{green_initial}
\end{eqnarray}
and we assume that at long times the Green's functions vanish.
Then one observes that Eq. (\ref{moments}) describes the relaxation of the initial equilibrium moments $\langle q^2\rangle=T_0/(m\omega^2)$ and $\langle \dot q^2\rangle=T_0/m$, corresponding to the initial temperature of the system $T_0$,  toward the new equilibrium values $\langle q^2\rangle=T/(m\omega^2)$ and $\langle \dot q^2\rangle=T/m$, corresponding to the temperature of the bath $T$.

From Eqs. (\ref{moments}) and (\ref{green_derivatives}) one finds
\begin{eqnarray}
    \frac{d^2}{dt^2} \langle q^2(t)\rangle=
    \frac{2\,T_0}{m\,\omega^2}
    \left\{
    \dot R^2+R\ddot R\right\}+
    \frac{2}{m} 
    \left(T_0-T\right)
    \left\{
    R^2+G\dot R\right\}+
    \frac{2\,T}{m}
    \left\{
    S R-\omega^2 G^2\right\}.
\end{eqnarray}
Then substituting to Eq. (\ref{W_total}) yields for the work the following result
\begin{eqnarray}
    \frac{W(\tau)}{T}&=&
    \left(\frac{T_0}{T}+1\right)
    \left(\frac{\omega_*}{\omega}\right)^2+
    \left[\left(\frac{\omega_*}{\omega}\right)^2+1\right]
    \left[
    \frac{T_0}{T}\,R^2(\tau)-S^2(\tau)
    \right]
    +\frac{T_0}{T\,\omega^2} R(\tau) \ddot R(\tau)+S(\tau)R(\tau)\nonumber\\
    &+&
    \left(\frac{T_0}{T}-1\right)
    \left\{G(\tau)\dot R(\tau)+\left[\left(\frac{\omega_*}{\omega}\right)^2+1\right] \omega^2 G^2(\tau)\right\}.
    \label{W_final}
\end{eqnarray}

Next, consider the internal energy of the system $U(t)$.
After disconnecting from the bath, i.e. for $t>\tau$,  it keeps  the value 
\begin{eqnarray}
    U(\tau)=\frac{1}{2}\,m\,\omega^2 \langle q^2(\tau)\rangle+\frac{1}{2}\,m\,\langle \dot q^2(\tau)\rangle.
\end{eqnarray}
Substituting here  Eq. (\ref{moments}) yields
\begin{eqnarray}
    \frac{U(\tau)}{T}=1+\left(\frac{T_0}{T}-\frac{1}{2}\right)\, R^2(\tau)
    -\frac{1}{2}\,S^2(\tau)+
\left(    
    \frac{T_0}{2\,T}-1\right)\, \omega^2 G^2(\tau)
    +\frac{T_0}{2\,T\,\omega^2}\, \dot R^2(\tau).
    \label{U}
\end{eqnarray}
The change of the internal energy of the system  during the process is $\Delta U(\tau)=U(\tau)-U(0^-)=U(\tau)-T_0$, and we find 
\begin{eqnarray}
    \frac{\Delta U(\tau)}{T}=1-\frac{T_0}{T}+\left(\frac{T_0}{T}-\frac{1}{2}\right)\, R^2(\tau)
    -\frac{1}{2}\,S^2(\tau)+
\left(    
    \frac{T_0}{2\,T}-1\right)\, \omega^2 G^2(\tau)
    +\frac{T_0}{2\,T\,\omega^2}\, \dot R^2(\tau).
    \label{DU}
\end{eqnarray}

Eqs. (\ref{W_final}) and (\ref{DU}) for $W$ and $\Delta U$ determine the energetics of the process for any values of the process duration $\tau$.  
Given $W$ and $\Delta U$, the energy change of the bath is determined by Eq. (\ref{law11}), 
$\Delta U_b=W-\Delta U$.  To make further progress, we need 
explicit expressions for the Green's functions. 
In Sec. V we shall consider 
a more specific model for which the Green's functions are known, 
and functions $W(\tau)$ and $\Delta U(\tau)$ can be explicitly evaluated. Meanwhile, in the next section we consider the limiting case $\tau\to\infty$, when only generic asymptotic  properties of the Green's functions are required.

To finish this section, let us note that in limit $\tau\to 0$ Eqs. (\ref{W_final}) and (\ref{DU})
give for $W(\tau)$ and $\Delta U(\tau)$ zero values, $W(0)=\Delta U(0)=0$. This result is physically plausible and can be directly verified taking into account 
the initial values of the Green's functions, Eq. (\ref{green_initial}), and their derivatives
\begin{eqnarray}
   \dot G(0)=1,\quad  \dot R(0)=\dot S(0)=0, \quad \ddot R(0)=-\omega^2-\omega_c^2=
    -\omega^2-2\omega_*^2.
    \label{green_initial2}
\end{eqnarray}
These relations follow from Eqs. (\ref{green_derivatives}), (\ref{green_initial}), (\ref{aux1A}), and (\ref{aux1AA}).

\section{Asymptotic results for $\tau\to\infty$}
The asymptotic properties of the Green's functions, which determine results (\ref{W_final}) and (\ref{DU}), are typically generic, but in some cases they may depend on a specific form of the dissipation kernel $K(t)$ and the system frequency $\omega$. 
For example, in the presence of localized modes the Green's functions at long time oscillate and their long time  limits do not exist~\cite{Dhar,Plyukhin}. 
In this paper we do not consider such cases and assume  
a more typical scenario when the system connected to a thermal bath eventually thermalizes. 
Thermalization implies that the Green's functions and their derivatives vanish at long times.  In that case, Eqs. (\ref{W_final}) and (\ref{DU}) in the limit $\tau\to\infty$ give for the work and the system's internal energy change the asymptotic values
\begin{eqnarray}
 W=
\left(\frac{\omega_*}{\omega}\right)^2
\left(T+T_0\right),\quad
\Delta U=T-T_0,
\label{work_asymp}
\end{eqnarray}
and the  asymptotic energy change of the bath  $\Delta U_b=W-\Delta U$ equals
\begin{eqnarray}
    \Delta U_b=\left(\frac{\omega_*}{\omega}\right)^2(T+T_0)-T+T_0.
    \label{DUb}
\end{eqnarray}
One observes that the asymptotic energy change for the system 
$\Delta U$ is in accordance with macroscopic thermodynamics:
It depends only on the temperature difference and 
does not depend on the coupling work.
For the bath the situation may be  different. Eventually, the work dissipates entirely into the bath and contributes 
to the change of the bath's internal energy $\Delta U_b$. As a result, 
the sign of $\Delta U_b$ may be determined not only by the temperature difference, but also by $W$.  

According to Eq. (\ref{work_asymp}),  the asymptotic work  $W$ is  positive and increases when the oscillator frequency decreases. It may be   negligible compared to $\Delta U$ and $\Delta U_b$, as the XFT assumes, only for a very high oscillator frequency,
\begin{eqnarray}
    \omega\gg\omega_*.
\end{eqnarray}
If that is not the case, the sign of the internal energy change of the bath $\Delta U_b$
may be anomalous from the point of view of a theory based on the weak coupling approximation. Indeed, one observes that 
for lower frequencies $\Delta U_b$ is positive for any temperature difference,
\begin{eqnarray}
    \Delta U_b>0 , \quad \text{for}\quad \omega\le\omega_*.
\end{eqnarray}
That is normal  when the system  is hotter than the bath ($T_0>T$), but anomalous  otherwise  ($T\ge T_0$).

The bath energy change  may be anomalous also for higher frequencies, 
$\omega>\omega_*$. 
In that case  Eq. (\ref{DUb})  
can be presented as
\begin{eqnarray}
    \Delta U_b=c\,(T_*-T), 
    \label{ax1}
\end{eqnarray}
with
\begin{eqnarray}
c=1-\left(\frac{\omega_*}{\omega}\right)^2,
\quad
T_*=\left(\frac{\omega^2+\omega_*^2}{\omega^2-\omega_*^2}\right)\,T_0>T_0.
\end{eqnarray}
According to Eq. (\ref{ax1}), the energy transfer to the bath 
appears to be 
normal, with the initial temperature of the system $T_0$ replaced by $T_*$: the bath gains energy when $T<T_*$ and losses energy when $T>T_*$. Since $T_*>T_0$,
the bath energy change  is anomalous when
$T_0\le T\le T_*$.
Note that for $\omega\gg \omega_*$, $T_*\to T_0$
and the temperature interval of the anomalous energy change for the bath 
shrinks to zero.

 Summarizing,  in the limit $\tau\to\infty$,
the work $W$ involved in the XFT protocol in general cannot be neglected, but it affects only the energy balance for the bath, and  not for the system.
The system energy change $\Delta U$ is always normal (positive for $T_0<T$ and negative for $T_0>T)$, whereas  
the bath energy change may be either normal or anomalous.
The conditions of anomalous energy balance for the bath are
\begin{eqnarray}
    T_0\le T, \quad \text{for}\quad \omega\le\omega_*,\nonumber\\
    T_0\le T\le T_*, \quad \text{for}\quad \omega>\omega_*.
    \label{anom_cond}
\end{eqnarray}
In that  cases the bath is hotter than (or has the same temperature as) the system, yet it 
gets an extra energy at the expense of the work  involved  in the process.  For the special case $\omega>\omega_*$ and $T=T_*$, the bath is hotter than the system, yet it neither lose nor get any energy, $\Delta U_b=0$. In that case the work goes entirely to increase the energy of the system, $\Delta U=W$.

Although the energy balance for the bath may be of interest from a theoretical point of view,  it is often immaterial in applications when one focuses on the system only. 
In view of the above findings, it is natural to ask whether
the energy balance may be anomalous not only for the bath (as we found for $\tau\to\infty$), 
but also for the system, if the process duration $\tau$ is finite. 
We shall see in the next section that the answer is affirmative.

\section{More specific model}
Explicit expressions for the Green's functions $G(t),R(t),S(t)$ depend on the memory kernel $K(t)$ in the Langevin equation (\ref{gle}).
The simplest and historically the first Langevin model is the Markovian one, which  corresponds to the delta-function ansatz
for the dissipation kernel
\begin{eqnarray}
    K(t)\to \gamma\, \delta(t), \quad \gamma=\int_0^\infty K(t)\,dt.
    \label{gamma}
\end{eqnarray}
While sufficient for many applications, the Markovian  approximation is clearly inappropriate for the problem at hand
because for the delta-like kernel 
the characteristic frequencies $\omega_c$ and $\omega_*=\omega_c^2/2$ diverge, see Eq. (\ref{K0}).
The kernel $K(t)$ is determined by the spectral density of the bath, and Markovian anzatz (\ref{gamma}) corresponds to the physically unrealistic 
assumption that the spectral density does not vanish in the high-frequency limit~\cite{Weiss}.

In this section we consider a non-Markovian  model where the dissipation kernel in the Langevin equation (\ref{gle}) 
has the specific form
\begin{eqnarray}
  K(t)=\frac{\mu\,\omega_0^2}{4}\,[J_0(\omega_0t)+J_2(\omega_0t)]
    =\frac{\mu\,\omega_0}{2}\,\frac{J_1(\omega_0 t)}{t}.
\label{K2}
\end{eqnarray}
Here $J_n(x)$'s are Bessel functions of the first kind, $\omega_0$ is a characteristic frequency of the model, and $\mu$ is a positive dimensionless parameter.
The kernel oscillates with an amplitude decaying with time as $t^{-3/2}$.
The squares of the characteristic frequencies are  
\begin{eqnarray}
    \omega_c^2=K(0)=\frac{\mu\,\omega_0^2}{4}, \quad 
\omega_*^2=\frac{\omega_c^2}{2}=\frac{\mu\,\omega_0^2}{8}.
\end{eqnarray}

The generalized Langevin equation with kernel (\ref{K2}) may correspond to many combinations of parameters of the  Caldeira-Leggett model.  In particular, it describes   the Rubin model where the system is the terminal isotope atom of mass $m$ in the  semi-infinite harmonic chain of atoms with the identical masses $m_i=m_0$~\cite{Weiss}. The chain plays the role of the bath. In our model, the isotope is also 
subjected to the additional harmonic potential $m\omega^2 q^2/2$.
For the Rubin model, 
the frequency $\omega_0$ has the meaning of the maximal frequency of the normal modes of the chain (the Debye frequency),
and $\mu$ is the  mass ratio, $\mu=m_0/m$.

For the given model Eq. (\ref{work_asymp}) for the asymptotic work (for $\tau\to\infty$) takes the form
\begin{eqnarray}
    W=\frac{\mu}{8}\,\left(\frac{\omega_0}{\omega}\right)^2\,(T+T_0).
    \label{work_limit2}
\end{eqnarray}
Analyzing  this and other results below, we can treat $(\mu,\omega_0,\omega)$ either as independent parameters, or interpret them in terms of the Rubin model. In the latter case, the parameters are not independent but connected by relations
\begin{eqnarray}
    \mu=m_0/m, \quad 
    \omega_0^2=4 k_0/m_0,\quad
    \omega^2=k/m,
\end{eqnarray}
where $k$ and $k_0$ are the stiffness coefficients for the oscillator's potential and for the lattice forming the bath, respectively. Then for the Rubin model the asymptotic work (\ref{work_limit2}) takes the form
\begin{eqnarray}
     W=\frac{k_0}{k}\,\frac{T+T_0}{2},
    \label{work_limit3}
\end{eqnarray}
which does not involve the masses of the system  and bath's particles. The asymptotic work is small, in accordance with the XFT's assumption,  for $k\gg k_0$, but that  appears to be a rather  unrealistic condition.

As was shown in Ref.~\cite{Plyukhin}, for the present model 
with $\mu<2$, there is  a critical value of the oscillator frequency 
\begin{eqnarray}
  \omega_e=\omega_0\,\sqrt{1-\mu/2},
  \label{omega_e}
\end{eqnarray}
(subscript $e$ stands for ``ergodic")  
which separates two types of the system's  behavior, namely 
with and without a localized mode.
For $\omega\le \omega_e$, there is no  localized modes, and the system thermalizes, i.e. reaches thermal equilibrium with the bath at long times. For $\omega>\omega_e$,  a localized mode develops, and the system, instead of thermalizing, evolves toward a cyclostationary state whose  parameters oscillates in time. 

In this paper we consider only the low-frequency case
$\omega\le \omega_e$, when the system thermalizes at long times. In that case  the Green's functions
can be expressed in the integral form as follows~\cite{Plyukhin}
 \begin{eqnarray}
 &&\!\!\!\!\!\!\!\!
 \omega_0G(t)\!=\!\frac{4\mu}{\pi}\!
   \int_0^1 \!
   \frac{\sin(x\,\omega_0\,t)\,x\,\sqrt{1-x^2}}
{D(x)}\,dx,\nonumber\\
&&\!\!\!\!\!\!\!\!
R(t)\!=\!\frac{4\mu}{\pi}\!
   \int_0^1 \!
   \frac{\cos(x\,\omega_0\,t)\,x^2\,\sqrt{1-x^2}}{D(x)}\,dx, \nonumber\\
&&\!\!\!\!\!\!\!\!
S(t)\!=\!1\!-\!\frac{4\lambda\mu}{\pi}\!
   \int_0^1 \!
   \frac{[1-\cos(x\,\omega_0\,t)]\,\sqrt{1-x^2}}{D(x)}\,dx, 
   \label{greens}
\end{eqnarray}
where the denominator of the integrands  is 
\begin{eqnarray}
D(x)=4(1\!-\!\mu) x^4\!+\![4\lambda (\mu\!-\!2)\!+\!\mu^2]\,x^2\!+\!4\lambda^2, 
\end{eqnarray}
and $\lambda$ denotes the square of the dimensionless oscillator frequency in units of $\omega_0$.
\begin{eqnarray}
\lambda=(\omega/\omega_0)^2.
\label{lambda}
\end{eqnarray}
The integral expressions (\ref{greens}) are reduced to closed forms only for two special cases $(\mu=1,\lambda=1/4)$ and $(\mu=1, \lambda=1/2)$; the latter is discussed below.
Note that  the condition of thermalization $\omega\le\omega_e$ in terms of $\lambda$ takes the form
\begin{eqnarray}
    \lambda\le \lambda_e=1-\mu/2.
\end{eqnarray}

For the present model, Eqs. (\ref{W_final}) and (\ref{DU}) for the work and the system's energy change become more concise when written in terms of $\lambda$ (rather than $\omega$):
\begin{eqnarray}
    \frac{W(\tau)}{T}&=&
    \left(\frac{T_0}{T}+1\right)
    \frac{\lambda_*}{\lambda}+
    \left[\frac{\lambda_*}{\lambda}+1\right]
    \left[
    \frac{T_0}{T}\,R^2(\tau)-S^2(\tau)
    \right]
    +\frac{T_0}{T\,\lambda}\, R(\tau) \left[\frac{\ddot R(\tau)}{\omega_0^2}\right]+S(\tau)R(\tau)\nonumber\\
    &+&
    \left(\frac{T_0}{T}-1\right)
    \left\{\omega_0G(\tau)\,\left[\frac{\dot R(\tau)}{\omega_0}\right]+\left[\frac{\lambda_*}{\lambda}+1\right] \lambda^2\, [\omega_0\,G(\tau)]^2\right\},
    \label{result1}\\
    \frac{\Delta U(\tau)}{T}&=&1-\frac{T_0}{T}+\left(\frac{T_0}{T}-\frac{1}{2}\right)\, R^2(\tau)
    -\frac{1}{2}\,S^2(\tau)+
\left(    
    \frac{T_0}{2\,T}-1\right)\, \lambda\,[\omega_0\,G(\tau)]^2
    +\frac{T_0}{2\,T\,\lambda}\, 
    \left[\frac{\dot R (\tau)}{\omega_0}\right]^2,
    \label{result2}
\end{eqnarray}
where $\lambda_*=(\omega_*/\omega_0)^2$. These
expressions also determine the internal energy of the bath 
\begin{eqnarray}
    \Delta U_b=W-\Delta U.
    \label{result3}
\end{eqnarray}
Note the functions $R$, $S$, $\omega_0G$, $\dot R/\omega_0$, and $\ddot R/\omega_0^2$ in the above expressions are dimensionless. 
Below we consider specific sets of parameters for which  Eqs. (\ref{result1})-(\ref{result3}) describe different scenarios.

\subsection{Special case $\mu=1, \lambda=1/2$: Work is non-negligible}
For this case
\begin{eqnarray}
\omega^2=\omega_0^2/2, \quad 
\omega_c^2=\omega_0^2/4, \quad
\omega_*^2=\omega_0^2/8,\quad
\lambda_*=(\omega_*/\omega_0)^2=1/8, \qquad \lambda_*/\lambda=1/4, 
\end{eqnarray}
and   expressions (\ref{greens}) for the Green's functions are reduced to compact forms in terms of the Bessel functions
\begin{eqnarray}
    \omega_0 G(t)=2\,J_1(\omega_0 t), \quad
    R(t)=J_0(\omega_0 t)-J_2(\omega_0 t), \quad
    S(t)=J_0(\omega_0t).
\end{eqnarray}
Substituting these expressions into Eqs. (\ref{result1})-(\ref{result3}) yields $W$, $\Delta U$, $\Delta U_b$ as functions of the process duration $\tau$
and the ratio of temperatures $T_0/T$.
The results are  presented in Fig. 1.  They show that for any $\tau$,  work $W$ is not small compared to $\Delta U$ and $\Delta U_b$, i. e. the assumption of the XFT is not satisfied.

\begin{figure}[t]
  \includegraphics[width=17cm]{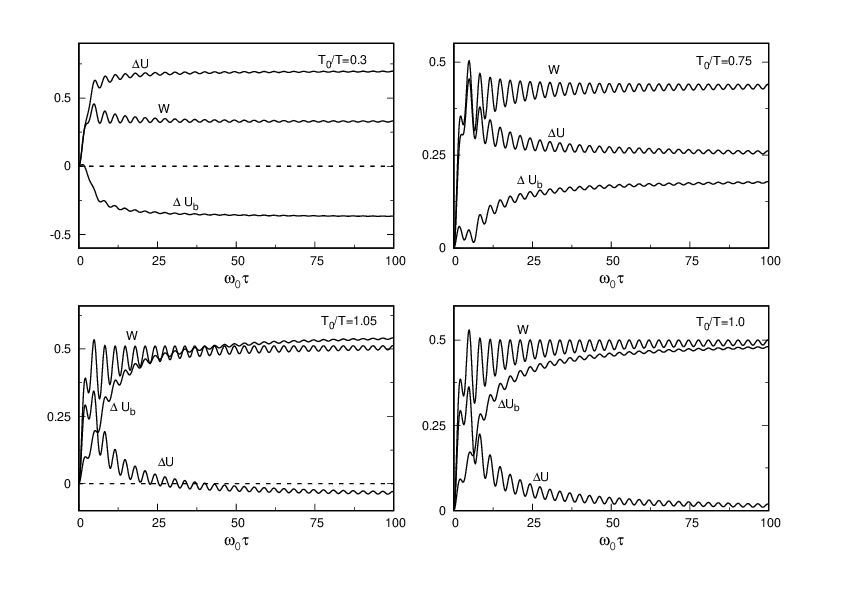}
\caption{Case $\mu=1, \lambda=1/2$. Work $W$, the internal energy changes of the system $\Delta U$ and the bath $\Delta U_b$ (in units of the temperature of the bath $T$)   as functions of the process duration $\tau$ for different ratios $T_0/T$ of the initial temperatures of the system $T_0$ and the temperature of the bath $T$.  
For all values of $\tau$ and $T_0/T$, work $W$ is comparable with $\Delta U$ and $\Delta U_b$, so the assumption of the XFT is not satisfied.
For $T_0/T\ge 1$ (plots at the bottom) the energy change of the system may be anomalous in the sense that  the system is initially hotter than the bath, yet it receives extra energy, $\Delta U>0$.}
\label{fig1}
\end{figure}

For processes with large $\tau$ the results corroborate  our findings in Sec. 4 for $\tau\to\infty$. The energy change of the system is normal, i.e $\Delta U\ge 0$ for $T_0/T\le 1$ and    $\Delta U\le 0$ for $T_0/T\ge 1$. 
The energy change of the bath $\Delta U_b$ may be either normal, or anomalous.
Note that for the present case $\omega>\omega_*$.  Then,  according to (\ref{anom_cond}), the energy change of the bath is anomalous when 
\begin{eqnarray}
    T_0\le T<T_*=\frac{\omega^2+\omega_*^2}{\omega^2-\omega_*^2}\,T_0=\frac{5}{3}\,T_0,
\end{eqnarray}
or
\begin{eqnarray}
    \frac{3}{5}<\frac{T_0}{T}\le 1.
\end{eqnarray}
This is illustrated  by the plots at the top of Fig. 1 for $T_0/T=0.3$ and $T_0/T=0.75$. In both cases the bath is hotter than the system, but  it loses energy only in the first case and gains energy in the second case.

The energy change of the system, which is often a more interesting quantity than that of the bath, 
is normal for $\tau\to\infty$, but may be anomalous for finite $\tau$. For $T_0/T=1.05$ (bottom left at Fig. 1) the system is slightly hotter than the bath, yet for $\tau\lesssim 30\,\omega_0$ it gains extra energy, $\Delta U>0$. The interval of $\tau$ corresponding to the anomalous energy change of the system increases when $T_0$ approaches $T$ from above and diverges in the limit $T_0/T\to 1^+$. For $T=T_0$ (bottom right at Fig. 1) the system receives energy for any finite $\tau$. The expectation of the standard thermodynamics that for $T_0=T$ the system gets no extra energy, $\Delta U=0$, is realized only in the limit $\tau\to\infty$.  


\subsection{Case $\mu=0.02,\,\lambda=0.5$: Work may be negligible}
 For this case, as for other cases discussed below,  the Green's functions cannot be reduced to  closed forms and 
 are given by Eq. (\ref{greens}). 
Substituting these expressions in Eqs. (\ref{result1})-(\ref{result3}) gives the results presented in Fig. 2.
As the two plots at top show, the work is small and  the assumption of the XFT is satisfied, provided  $\tau$ is not too small and the ratio of temperatures $T_0/T$ is not too close to $1$. 

\begin{figure}[t]
 \includegraphics[width=17cm]{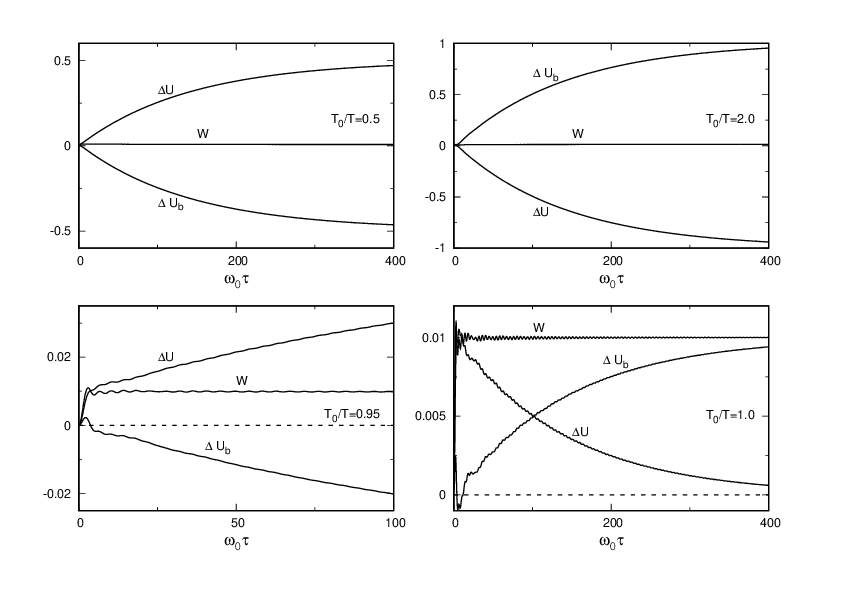}
 \caption{Case $\mu=0.02, \lambda=0.5$. Work $W$ can be negligible compared to the internal energy changes of the system $\Delta U$ and the bath $\Delta U_b$ (two plots at top), unless  the ratio of temperatures is close or equal to $1$ (two plots at bottom).}
\label{fig2}
\end{figure}

When  $T_0/T$ is getting close to $1$,  work $W$ may be non-negligible for finite $\tau$, see the figure at bottom-left.
For the isothermal process, $T_0/T=1$, the work is larger than both 
$\Delta U$ and $\Delta U_b$ for any $\tau$, see the plot at bottom right.

The result that $\Delta U$ and $\Delta U_b$ are non-zero for the isothermal process, $T=T_0$, may appear to contradict
conventional thermodynamics which asserts that no heat exchange occurs when the temperatures of the systems are the same.
Actually, there is of course no contradiction because for the considered process $\Delta U$ and $\Delta U_b$ cannot be qualified as 
heat but rather originate from work performed by an external agent. 
The energy change of the system $\Delta U$ approaches zero only in the limit  $\tau\to\infty$, because  the energy  the system obtained from work eventually dissipates into the bath.

If we interpret results in terms of the Rubin model, then the parameters are related as follows:
\begin{eqnarray}
    \lambda=\left(
    \frac{\omega}{\omega_0}
    \right)^2=\frac{m_0}{m}\,\frac{k}{k_0}=\mu\,\frac{k}{k_0}.
\end{eqnarray}
For the present case $k/k_0=\lambda/\mu=25$, i. e. $k\gg k_0$. For a lattice-like bath,  this relation may be  unrealistic since the potential applied to the system is often much softer than that of the lattice, $k\ll k_0$.

\subsection{Case $\mu=0.02,\,\lambda=0.002$: Heavy particle in a soft potential}
For this case, analyzed in terms of the Rubin model, one gets
\begin{eqnarray}
    \frac{k}{k_0}=\frac{\lambda}{\mu}=0.1.
\end{eqnarray}
This setting appears to be more realistic compared to the one discussed above. It describes a  Brownian particle, which is much heavier than the particles of the lattice bath, subjected to the external harmonic potential which is much softer than that of the lattice.   The results, shown in Fig. 3, suggest that in this and similar cases the work is in general non-negligible and the assumption of the XFT is not satisfied.

\begin{figure}[t]
 \includegraphics[width=17cm]{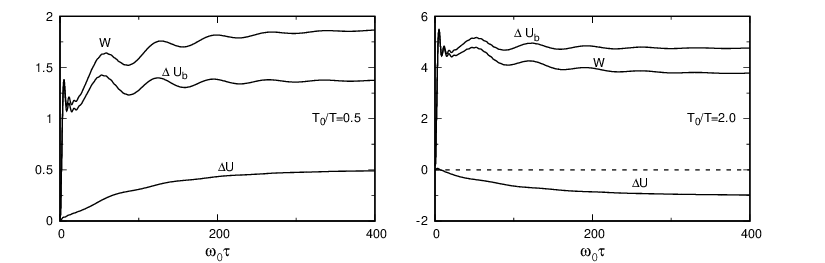}
 \caption{Case $\mu=0.02, \lambda=0.002$. In terms of the Rubin model the case corresponds to a heavy system, $m\gg m_0$ subjected to a soft potential, $k\ll k_0$. }
\label{fig3}
\end{figure}

\subsection{Case $\mu=0.1,\,\lambda=0.9, T_0/T=5$: Negative work}
For all settings considered above the work was positive, i.e. the connection and subsequent disconnection of the system and bath requires energy. This feature turns out to be not general. Fig. 4 shows that the work may be  negative for a finite range  of $\tau$.  
A negative work means that the process operates as a heat engine, doing work on the surroundings.

\begin{figure}[t]
 \includegraphics[width=17cm]{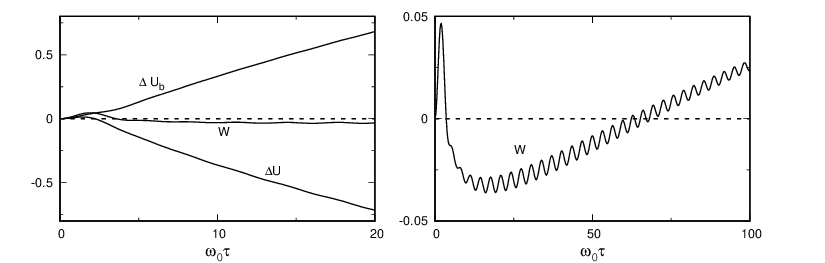}
 \caption{Case $\mu=0.1, \lambda=0.9, T_0/T=5$. Left: the work $W$ and the changes of internal energies of the system $\Delta U$ and the bath $\Delta U_b$ on the smaller scale of the process duration $\tau$. Right: The work on the larger scale  of $\tau$.  On the even larger scale of $\tau$ the work reaches the limit described by Eq. (\ref{work_limit2}).  }
\label{fig4}
\end{figure}

\section{Conclusion}
The chief motivation of this study is to verify whether the XFT's assumption of a negligible  work is valid when one of  the systems involved in the process is microscopic. 
We focused on the model  where the small system is a single classical oscillator and the large system is  a thermal bath composed of many independent oscillators. We found that, although there is a range of parameters for which the assumption of a negligible work is valid, it is rather restrictive and in general the assumption does not hold.

As a result, the energetics of the process with finite duration $\tau$ may be anomalous from the point of view of macroscopic thermodynamics and differ from the prediction of the XFT. In particular, the initially hotter system
may permanently receive extra energy. This scenario does not contradict the second law because the process  is not spontaneous but involves an external (positive) work.

Besides the finding that the work in general is non-negligible, we also observed that for a finite process duration the work's sign may be positive or negative. In the latter case the process addressed by the XFT may operate as a heat engine. This perhaps is a remarkable feature considering that,  unlike the majority of other heat engines, the process does not involve variation of any parameters, except the one describing the system-bath coupling.

As a more general remark, for a strong system-bath coupling there appears to be no unambiguous  way to define heat. (This is somewhat unfortunate for a topic referred to as ``heat exchange".) Instead of the set of thermodynamic variables
$(W, \Delta U, Q)$,
it may be  more relevant to use, as we did in this paper,  the set of mechanical variables $(W, \Delta U, \Delta U_b)$, replacing heat $Q$ by the change of the internal energy of the bath $\Delta U_b$. 

\vspace{0.5cm}
\noindent
{\it Data Availability:} Row data sets generated during this study are
available upon request.

\vspace{0.5cm}
\noindent
{\it Competing Interests:} The author has no competing interests to declare.

\renewcommand{\theequation}{A\arabic{equation}}
  \setcounter{equation}{0}  

  \section*{APPENDIX}  
  In this Appendix we evaluate the second moments
  $\langle q^2(t)\rangle$ and $\langle \dot q^2(t)\rangle$
governed by the Langevin equation (\ref{gle}). Solving that equation
using Laplace transforms,  one finds 
\begin{eqnarray}
    q(t)=q\, R(t)+\dot q\, G(t)+\frac{1}{m}(\xi*G)(t), \quad
    \dot q(t)=q\, \dot R(t)+\dot q\, R(t)+\frac{1}{m}(\xi*R)(t)
    \label{Q}
\end{eqnarray}
Here $q=q(0), \dot q=\dot q(0)$, the Green's functions $G(t)$ and $R(t)$ are determined in the Laplace domain
by relations
\begin{eqnarray}
    \tilde G(s)=\int_0^\infty\!\! e^{-st}G(t)\, dt=\frac{1}{s^2+s\tilde K(s) +\omega^2},\quad
    \tilde R(s)=s\,\tilde G(s),
\end{eqnarray}
and the symbol $*$ stands for the convolution, 
\begin{eqnarray}
    (\xi*G)(t)=\int_0^t G(t')\,\xi(t-t')\,dt', 
    \quad 
    (\xi*R)(t)=\int_0^t R(t')\,\xi(t-t')\,dt'.
\end{eqnarray}
It is also convenient to use the third Green's function  $S(t)$ 
defined in the Laplace domain as
\begin{eqnarray}
    \tilde S(s)=[s+\tilde K(s)]\,\tilde G(s)=\frac{1}{s}[1-\omega^2 \tilde G(s)].
\end{eqnarray}
In the time domain the Green's functions have initial values
\begin{eqnarray}
    G(0)=0, \quad R(0)=S(0)=1,
    \label{green_initialA}
\end{eqnarray}
and are connected by relations
\begin{eqnarray}
    \dot G(t)=R(t), \quad \dot S(t)=-\omega^2 G(t).
    \label{green_derivativesA}
\end{eqnarray}
Also, one can verify the validity of the relations
\begin{eqnarray}
   s\,\tilde G(s)=\tilde S(s)-\tilde K(s) \,\tilde G(s),\quad
   s\tilde R(s)-1=-\omega^2 \tilde G(s)-\tilde K(s) \tilde R(s),
   \label{green_relations}
\end{eqnarray}
which in the time domain have the forms
\begin{eqnarray}
    \dot G(t)=S(t)-(K*G)(t), \quad \dot R(t)=-\omega^2 G(t)-(K*R)(t).
    \label{aux1A}
\end{eqnarray}
Here the second equation can be obtained by differentiating the first one. 
Differentiating the second equation yields another useful relation
\begin{eqnarray}
    \ddot R(t)=-\omega^2 R(t)-K(t)-(K*\dot R)(t).
    \label{aux1AA}
\end{eqnarray}

Squaring and averaging  Eq. (\ref{Q})  one gets
\begin{eqnarray}
    \langle q^2(t)\rangle&
    =&\frac{T_0}{m\,\omega^2}\,R^2(t)+\frac{T_0}{m}\, G^2(t)+\frac{1}{m^2}\,\langle (\xi*G)^2(t)\rangle,\nonumber\\
    \langle \dot q^2(t)\rangle&=&\frac{T_0}{m\,\omega^2}\,\dot R^2(t)+\frac{T_0}{m}\, R^2(t)+\frac{1}{m^2}\,\langle (\xi*R)^2(t)\rangle.
    \label{aux2A}
\end{eqnarray}
The last terms in these expressions can be evaluated  replacing the integration over the square 
$(0,t)\times (0,t)$ by two times the integral over a half-square triangle. 
Then, taking into account the fluctuation-dissipation relation (\ref{fdr}) and Eq. (\ref{aux1A}),
one gets
\begin{eqnarray}
    \langle (\xi*G)^2(t)\rangle=\frac{m\,T}{\omega^2}\,[1-S^2(t)]-
    m\,T\,G^2(t), \quad
\langle (\xi*R)^2(t)\rangle=m\,T\,[1-R^2(t)-\omega^2 G^2(t)].
\end{eqnarray}
Substituting into Eq. (\ref{aux2A}) yields
\begin{eqnarray}
    \langle q^2(t)\rangle&=&\frac{T_0}{m\,\omega^2}\,R^2(t)+
    \frac{1}{m}\left(T_0-T\right)
    \,G^2(t)+
    \frac{T}{m\,\omega^2}\, [1-S^2(t)],\nonumber\\
    \langle \dot q^2(t)\rangle&=&\frac{T_0}{m\,\omega^2}\,\dot R^2(t)+
\frac{T_0}{m}\,R^2(t)+
    \frac{T}{m}\, [1-R^2(t)-\omega^2 G^2(t)].
\end{eqnarray}
These are expressions in Eq. (\ref{moments}) of the main text.


\end{document}